\documentclass[twocolumn,showpacs,preprintnumbers]{revtex4}
\usepackage{amsmath}
\usepackage{graphicx}
\usepackage{dcolumn}
\usepackage{bm}
\usepackage{subfigure}
\usepackage{times}
\begin{document}
%%%%%%%%%%%%%%%%%%%%%%%%%%%%%%%%%%%%%%%%%%%%%%%%%%%%%%%%%%%%%%%%%%%%%%%%%%%%%%%%%%%%%%%%%%%%%
\title{Quantum state engineering in a cavity by Stark chirped rapid adiabatic passage }
\author{M. Amniat-Talab$^{1}$}
\email{amniyatm@u-bourgogne.fr}
\author{R. Khoda-Bakhsh$^{1}$}
\author{S. Gu\'{e}rin$^{2}$}
\email{sguerin@u-bourgogne.fr}
 \affiliation{$^{1}$Physics Department, Faculty of Sciences, Urmia
University, P.B. 165, Urmia, Iran.\\
$^{2}$Laboratoire de Physique, UMR CNRS 5027, Universit\'{e} de
Bourgogne, B.P. 47870, F-21078 Dijon, France.}
%%%%%%%%%%%%%%%%%%%%%%%%%%%%%%%%%%%%%%%%%%%%%%%%%%%%%%%%%%
\date{\today }
%%%%%%%%%%%%%%%%%%%%%%%%%%%%%%%%%%%%%%%%%%%%%%%%%%%%%%%%%%
\begin{abstract}
We propose a robust scheme to generate single-photon Fock states
and  atom-photon and atom-atom entanglement in atom-cavity
systems. We also present a scheme for quantum networking between
two cavity nodes using an atomic channel. The mechanism is based
on Stark-chirped rapid adiabatic passage (SCRAP) and half-SCRAP
processes in a microwave cavity. The engineering of these states
depends on the design of the adiabatic dynamics through the static
and dynamic Stark shifts.
\end{abstract}
%%%%%%%%%%%%%%%%%%%%%%%%%%%%%%%%%%%%%%%%%%%%%%%%%%%%%%%%%%
\pacs{42.50.Dv, 03.65.Ud, 03.67.Mn, 32.80.Qk }
 \maketitle
%%%%%%%%%%%%%%%%%%%%%%%%%%%%%%%%%%%%%%%%%%%%%%%%%%%%%%%%%
\section{Introduction\label{intro}}
Quantum-state engineering (QSE), i.e., active control over the
coherent dynamics of suitable quantum-mechanical systems to
achieve a preselected state (e.g. entangled states or multi-photon
field states) of the system, has become a fascinating prospect of
modern physics. With concepts developed  in atomic and molecular
physics, the field has been stimulated further by the perspectives
of quantum computation and communication.

Single-photon states act as flying qubits  in quantum cryptography
\cite{jennewein,naik,tittel}. Key bits can be encoded on the
polarization states of single photons; the security of the
transmission is based on the fact that a single photon is
indivisible and its unknown quantum state cannot be copied. If
Alice sends to Bob  pulses containing more than one photon, it is
possible for a potential evesdropper  to measure the photon number
in each pulse without disturbing any photons. She can then split
off one photon from the pulses that originally contain more than
one photon, and retain this photon until Alice and Bob discuss
their bases. Such pulses are vulnerable to a photon-number
splitting attack. At this point, she can correctly measure the
photon polarization and obtain the corresponding bit value without
creating any error on Bob's side. Hence, implementation of quantum
cryptography would be secured by a source that emits only one
photon at a time. These states can also be used in the
error-tolerant quantum computing proposal of Gottesman and Chuang
\cite{Gottesman} which requires that a quantum resource be
supplied \emph{on demand}. Photon fields with fixed photon numbers
are also interesting from the point of view of fundamental physics
since they represent the ultimate non-classical limit of
radiation.

In the context of cavity quantum electrodynamics (CQED),
single-photon Fock states have been produced by  $\pi$-pulse
technique in a microwave cavity \cite{MaitrePRL97,varcoewalther}
and by the stimulated Raman adiabatic passage (STIRAP) technique
in an optical cavity \cite{hennrichPRL2000} based on the scheme
proposed in \cite{ParkinsPRL93} where the Stokes pulse is replaced
by a mode of a high-Q cavity. However the $\pi$-pulse technique is
not robust with respect to variations of pulse parameters and to
exact-resonance condition. The tripod STIRAP process has also been
studied in a system of a four-level atom interacting with a cavity
mode and two laser pulses, with a coupling scheme which has two
globally degenerate dark states \cite{gong}. The fractional STIRAP
process has also been studied in an optical cavity to prepare
atom-photon and atom-atom entanglement \cite{AmniatPRA05}.
Bichromatic adiabatic passage in a microwave cavity is another
robust technique that allows one to generate a controlled number
of photons in the cavity mode \cite{AmniatPRA04}.
 SCRAP is a
powerful technique in which the energy of a target state
$|+\rangle$ is swept through resonance by a slowly varying dynamic
Stark shift to induce complete population transfer from the ground
state $|-\rangle$ to the excited state $|+\rangle$. This method
uses (i) a laser pulse (the pump) tuned slightly away from the
one-photon resonance between the states, and (ii) a relatively
intense far-off-resonance pulse (the Stark field) that sweeps the
states through the resonance by inducing a dynamic Stark shift. A
time delay between the Stark  and  pump pulses, ensures that the
entire population is in the excited state at the end of the
process. In this paper we propose a robust scheme in the context
of CQED to generate single-photon, atom-photon and atom-atom
entangled states based on the SCRAP and the half-SCRAP techniques
proposed in the context of laser-driven systems
\cite{YatsenkoPRA99,YatsenkoOC02}.

One can transfer a quantum state by quantum networking. The basic
idea behind a quantum network is to transfer a quantum state from
one node to another node with the help of a carrier (a quantum
channel) such that it arrives intact. In between, one has to
perform a process of quantum  state transfer (QST) to transfer the
state from one node to the carrier and again from the carrier to
the destination node. In this paper, using the SCRAP technique in
a cavity, we propose a robust scheme for QST to transfer the
unknown state of a two-level atom to another atom where the atoms
are not directly interacting with each other. We also extend our
idea of QST to a quantum network, where we transfer the state of
one cavity to another spatially separated cavity. For this we use
long-lived atoms as carrier, and make use of the QST process to
transfer the state of the cavity to an atom and again to the
target cavity.

\section{construction of the effective Hamiltonian\label{model}}
\begin{figure}
\centerline{\subfigure{\includegraphics[width=4cm]{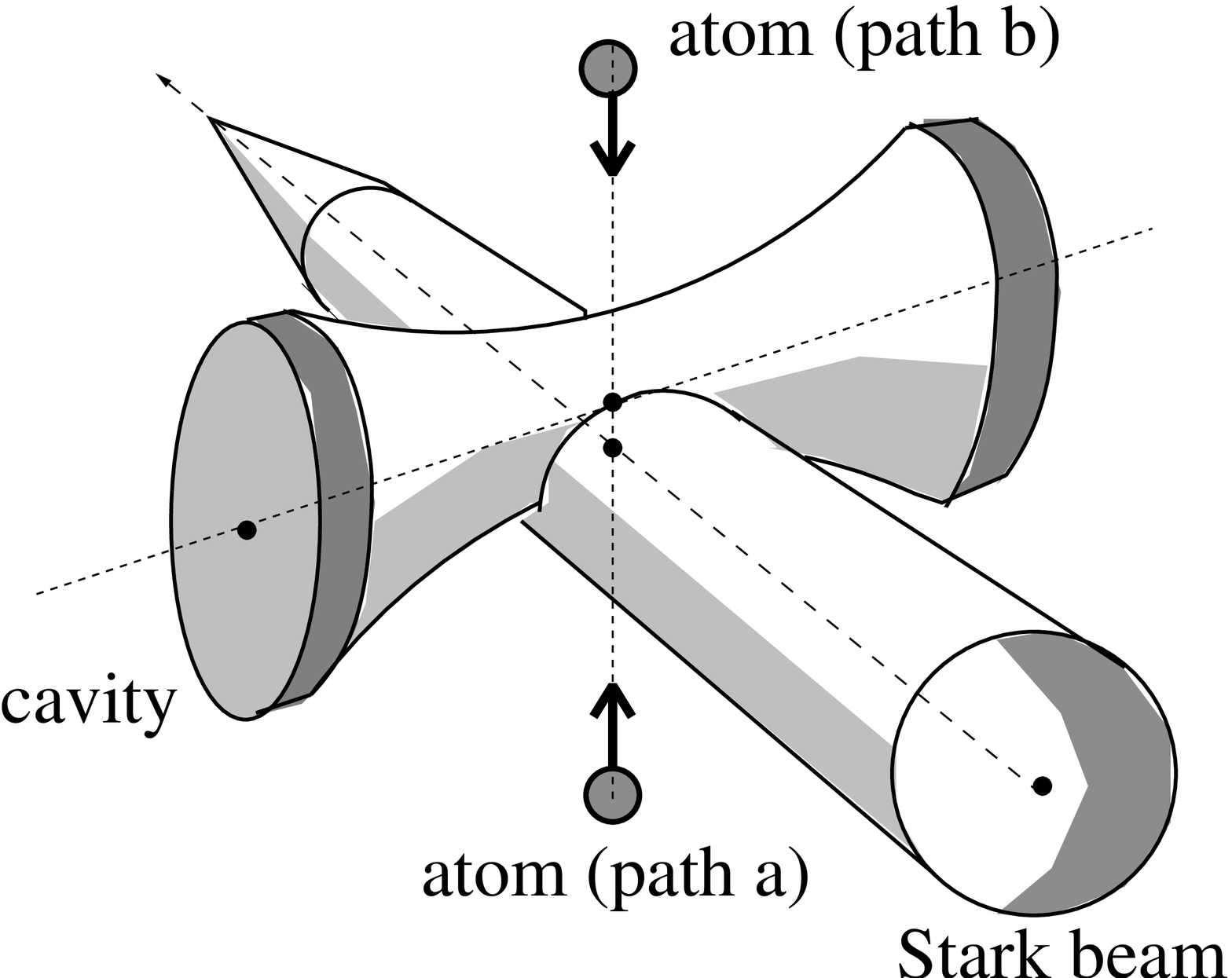}}
\subfigure{
\includegraphics[width=2.5cm]{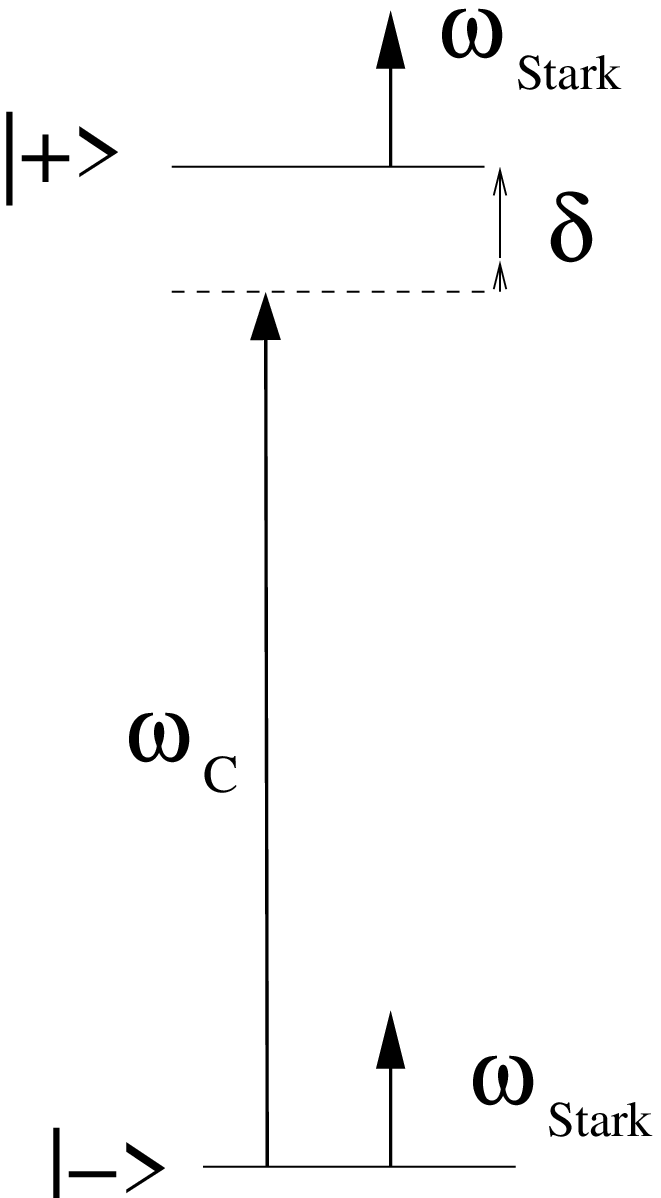}}}
 \caption{Experimental configuration and  linkage pattern of the system. Paths (a) and (b)
 refer to the paths drawn in Fig. \ref{fig2}.}\label{fig1}
\end{figure}

We consider a two-level atom of upper and lower states $|+\rangle$
and  $|-\rangle$ and of energy difference $E_{+}-E_{-}=\omega_{0}$
as represented in Fig. 1. We use atomic units in which $\hbar=1$.
The atom initially prepared in its upper state falls through a
high-Q microwave cavity with velocity $v$. The atom first
encounters the vacuum mode of the cavity with frequency
$\omega_{C}$ and waist $W_{C}$ and then the maser beam (Stark
field) with frequency $\omega_{S}$ and waist $W_{S}$. The cavity
field is near-resonant with the atomic transition, while the Stark
field is far-off-resonance  that sweeps the states through the
resonance by inducing a dynamic Stark shift. The distance between
the crossing points of the cavity and the maser axis with the
atomic trajectory is $d$. The travelling atom encounters time
dependent and delayed Rabi frequency of the cavity mode and the
Stark shift:
\begin{subequations}
\begin{eqnarray}
    G(t)&=&G_{0}~e^{-(\frac{vt}{W_{C}})^{2}}\label{Grabi},\\
    S(t)&=&S_{0}~e^{-(\frac{vt\pm d}{W_{S}})^{2}}\label{Srabi},
\end{eqnarray}
\end{subequations} where $S_{0}$ is  the peak value of the Stark
shift, and $G_{0}=\mu
    \sqrt{\frac{\omega_{C}}{2\epsilon_{0}V_{\text{mode}}}}$ is  the peak value of the cavity's Rabi frequency with  $\mu$,
    $V_{\text{mode}}$ respectively
    the dipole moment of  the atomic transition, and the
effective volume of the cavity mode. In Eq. \ref{Srabi}, the signs
$(+,-)$ correspond respectively to paths (a,b) in Fig. \ref{fig1}.

%%%%%%%%%%%%%%%%%%%%%%%%%%%%%%%%%%%%%%%%%%%%%%%%%%%%%%%%%%%%%%%%%%%%%%%%%%%%%
%%%%%%%%%%%%%%%%%%%%%%%%%%%%%%%%%%%%%%%%%%%%%%%%%%%%%%%%%%%%%%%%%%%%%%%%%%%%%%%%%%%%%%%%%%%%%%%%%%%%%%%%%%%%%%%%%%%%%
The detuning of the cavity mode from the atomic transition is
$\delta=\omega_{0}-\omega_{C}$. We take the frequency of the
cavity field such that $\delta$ is positive and very small with
respect  to $\omega_{C}$ and $\omega_{0}$. Also we assume
\begin{equation}\label{rabicondition}
    \text{max}\{|G(t)|\}\ll \omega_{0},\omega_{C}.
\end{equation}
Under these conditions, the counter-rotating terms can be
discarded in the rotating-wave approximation (RWA). The
semiclassical Hamiltonian of the atom-maser-cavity system can thus
be written as
\begin{eqnarray}\label{H}
H(t) &=& \omega_{C}a^{\dag}a
+\left(%
\begin{array}{cc}
  \omega_{0}+S_{+}(t) & 0 \\
  0 & S_{-}(t) \\
\end{array}%
\right)\nonumber\\
    &+&G(t)\left(%
\begin{array}{cc}
  0 & a \\
  a^{\dag} & 0 \\
\end{array}%
\right),
\end{eqnarray}
where $a,a^{\dag}$ are the annihilation and creation operators
 of the cavity mode,  and $S_{+,-}(t)$ are the dynamic Stark shifts (proportional
to the Stark field intensity) of the bare atomic states.  The
energy of the lower atomic state has been taken as $0$. This
Hamiltonian acts on the Hilbert space $\mathcal{H} \otimes
\mathcal{F}$ where $\mathcal{H}$ is the Hilbert space of the atom
generated by $|\pm\rangle$ and $\mathcal{F}$ is the Fock space of
the cavity mode generated by the orthonormal basis
$\{|n\rangle~;~n=0,1,2,\cdots\}$ with $n$ the photon number of the
cavity field.

The Hamiltonian $H(t)$ is block-diagonal in the subspaces $%
\{|+,n\rangle ,|-,n+1\rangle ;~n=0,1,2,...\}$, $|+,n\rangle \equiv
|+\rangle
\otimes |n\rangle $ and $|n\rangle $ is a $n$-photon Fock state. The vector $%
|-,0\rangle $ is not coupled to any other ones, i.e., $|-,0\rangle
$ is a stationary state of the system. One can thus restrict the
problem to the projection of the Hamiltonian in the subspace
$\mathcal{S}=\{|+,0\rangle ,|-,1\rangle \}$:
\begin{subequations}
\begin{eqnarray}
H_{P} &:=&PHP, \\
P &=&|+,0\rangle \left\langle +,0\right| +|-,1\rangle
\left\langle-,1\right|,
\end{eqnarray}
\end{subequations}
 if one considers the initial state $|+,0\rangle$. The effective
 Hamiltonian of the system in the subspace $\mathcal{S}$ can be
 written as
 \begin{equation}\label{Heff}
H^{\text{eff}}(t)=\left(
\begin{array}{cc}
\delta+S(t) & G(t) \\
 G(t) &  0
\end{array}%
\right),
\end{equation}%
where $S(t)=S_{+}(t)-S_{-}(t)$ is the difference of the Stark
shifts of the bare states.  The detuning
\begin{equation}\label{Delta}
    \Delta(t)=\delta+S(t),
\end{equation}
 is the effective  dynamic detuning
from one-photon resonance.

We remark that the Stark shifts of the ground and excited states
are different. The frequency $\omega_{\text{Stark}}$ of the Stark
field should be chosen such that it is not in resonance involving
the considered levels. Usually in atoms, one chooses  the carrier
frequency of the Stark field  much smaller than $\omega_{0}$ to
prevent e.g. the ionization effects. Under this condition, the
coupling of the state $|+\rangle$ to the other atomic upper states
(not included in the linkage pattern of  Fig. \ref{fig1}) induced
by the Stark field,  will be larger than their  coupling with the
state $|-\rangle$  (since it is farther  from them) giving
$|S_{+}(t)|\gg|S_{-}(t)|$. This results in a reduction of the
energy difference between the two states $|-\rangle$ and
$|+\rangle$, i.e., $S_{+}(t)<0$.

\section{Topology of the dressed eigenenergy surfaces\label{topology}}
\begin{figure}
\begin{center}
  \includegraphics[width=6cm]{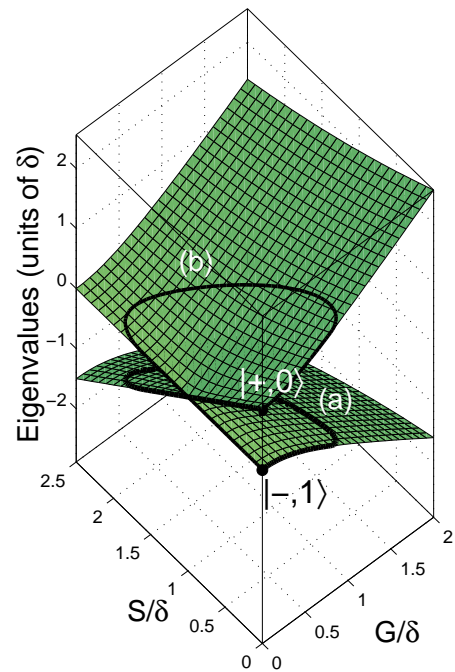}\\
  \caption{(Color online). Eigenenergy surfaces (in units of $\delta$) of $H^{\text{eff}}$ as functions of the cavity Rabi frequency
  $G$ and of the Stark shift $S$.  The solid paths correspond to  adiabatic
evolutions which start from $|+,0\rangle$ state and end at
$|-,1\rangle$.
   }\label{fig2}
   \end{center}
\end{figure}

The SCRAP process can be completely described
\cite{GuerinPRA01,GuerinPRA02} by the diagram of the two surfaces
\begin{equation}\label{scrap-E}
    E_{\pm}(G,\Delta)=\frac{1}{2}(\Delta\pm\sqrt{\Delta^{2}+4G^{2}}),
\end{equation}
which represent the eigenenergies as functions of the
instantaneous effective Rabi frequency $G$ and Stark shift $S$
(see Fig. \ref{fig2}). All the quantities are normalized with
respect to the static detuning $\delta$. The topology of these
surfaces, determined by the conical intersections, presents
insight into the various adiabatic dynamics which leads to
transfer a single photon into the cavity mode by designing
appropriate paths connecting the initial and  the chosen final
state. Each path corresponds to a choice of the envelope of the
pulses. In the adiabatic limit, when the pulses vary sufficiently
slowly,  the solution of the time-dependent  Schr\"{o}dinger
equation follows the instantaneous  eigenvectors, following the
path on the surface that is continuously connected to the initial
state. We start with the  state $|+,0\rangle$, i.e., the upper
atomic state with zero photons in the cavity field. Its energy is
shown in Fig. \ref{fig2} as the starting point of the  paths (a),
and (b). The paths shown in Fig. \ref{fig2} describe accurately
the dynamics if the time dependence of the envelopes is slow
enough according to the Landau-Zener \cite{L,Z} and
Dykhne-Davis-Pechukas \cite{D,DP} analysis. If two (uncoupled)
eigenvalues cross, the adiabatic theorem of Born and Fock
\cite{BF} shows that the dynamics follows diabatically the
crossing. This implies that the various dynamics shown in Fig.
\ref{fig2} are a combination of a global adiabatic passage around
the conical intersection and local diabatic evolutions through (or
in the neighborhood) of conical intersection of the eigenenergy
surfaces \cite{GuerinPRA01}.

 The eigenenergy surfaces display a conical intersection for $G=0$ and
$|S|=\delta$. In the plane $G=0$, the states $|+,0\rangle$ and
$|-,1\rangle$ do not interact. Figure \ref{fig2} shows two
possible paths corresponding to SCRAP with $\delta=+|\delta|$
which adiabatically connect the initial state $|+,0\rangle$ to the
final state $|-,1\rangle$. For the path (a), while the cavity
pulse is off, the Stark pulse is switched on and induces a
negative Stark shift $S(t)<0$.
%%%%%%%%%%%%%%%%%%%%%%%%%%%%%%%%%%%%%%%%%%%%%%%%%%%%%%%%%%%%%%%%%%%%%%%%%%%%%%%%%%%%%%%%%%%%%%%%%%%%
%%%%%%%%%%%%%%%%%%%%%%%%%%%%%%%%%%%%%%%%%%%%%%%%%%%%%%%%%%%%%%%%%%%%%%%%%%%%%%%%%%%%%%%%%%%%%
Thus, the Stark pulse makes the eigenstates get closer, and
induces a resonance with the cavity frequency. This resonance is
mute since the cavity pulse is still off, which results in the
true crossing in the diagram. The cavity pulse is switched on
after the crossing. Later the Stark pulse decreases while the
cavity pulse is still on. Finally, the cavity pulse is switched
off. The path (b) leads exactly to the same final effect: the
cavity pulse is switched on first (making the eigenvalues repel
each other as shown in the figure) before the Stark pulse $S(t)$,
which is switched off after the cavity pulse.

Inspection of the eigenenergy surfaces in Fig. \ref{fig2} shows
that an essential condition for paths (a),(b) is $S_{0}>\delta$
such that the dynamics goes through the conical intersection (on
the $G=0$ plane) between the upper surface (connected to the
initial state $|+,0\rangle$) and the lower surface (connected to
the final state $|-,1\rangle$). The crossing of this intersection
as $S$ varies with $G=0$, brings the  system into the lower
eigenenergy surface.

%%%%%%%%%%%%%%%%%%%%%%%%%%%%%%%%%%%%%%%%%%%%%%%%%%%%%%%%%%%%%%%%%%%%%%%%%%%%%
\begin{figure}
\includegraphics[width=6cm]{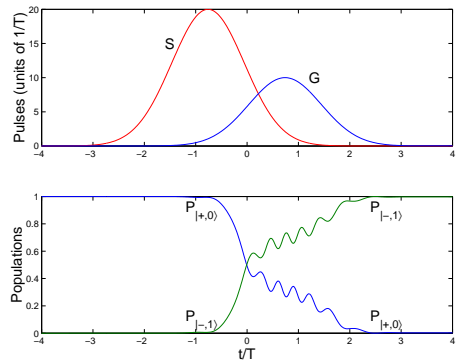}
 \caption{(Color online). Upper panel: the cavity-Stark sequence  of  Gaussian-shaped pulses
 [see Eqs. (\ref{Grabi}) and (\ref{Srabi})] with the same pulse duration $T$ and pulse parameters $S_{0}=20/T,~G_{0}=10/T,~\tau=1.5T$, and
 $\delta=10/T$.
  Lower panel:
 the  populations of the states $|+,0\rangle$ and $|-,1\rangle$ as functions of time which
   corresponds to the path (a) in Fig. \ref{fig2}.}\label{fig3}
\end{figure}
%%%%%%%%%%%%%%%%%%%%%%%%%%%%%%%%%%%%%%%%%%%%%%%%%%%%%%%%%%%%%%%%%%%%%%%%%%%%%%%
%%%%%%%%%%%%%%%%%%%%%%%%%%%%%%%%%%%%%%%%%%%%%%%%%%%%%%%%%%%%%%%%%%%%%%%%%%%%%
\begin{figure}
\includegraphics[width=6cm]{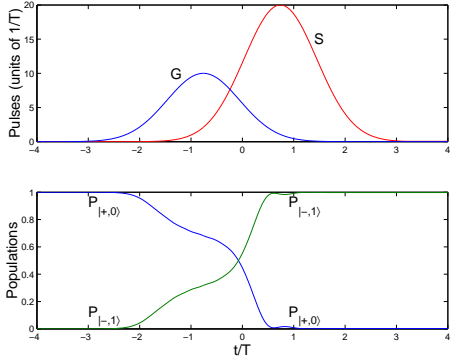}
 \caption{(Color online). Same  as Fig. \ref{fig3},
 but with the reverse pulse sequence of cavity-Stark which corresponds to the
path (b) in Fig. \ref{fig2}. The lower panel shows that the final
state of the system will be $|-,1\rangle$ as in Fig.
\ref{fig3}.}\label{fig4}
\end{figure}
%%%%%%%%%%%%%%%%%%%%%%%%%%%%%%%%%%%%%%%%%%%%%%%%%%%%%%%%%%%%%%%%%%%%%%%%%%%%%%%
\section{\label{numsim}numerical simulation}
The dynamics of the system is governed by the Schr\"{o}dinger
equation $i(\partial/\partial t)|\Phi (t)\rangle
=H^{\text{eff}}(t)|\Phi (t)\rangle $. The time dependence of
parameters of the system are Gaussians of the form (\ref{Grabi}),
(\ref{Srabi}) with the time delay $\tau=d/v$. Figure
\ref{fig3}(upper panel) shows the Stark and cavity Gaussian-shaped
pulses  with the pulse sequence of Stark-cavity and the same pulse
duration $T_{C}=W_{C}/v=T=T_{S}=W_{S}/v$ corresponding to the path
(a) in Fig. \ref{fig2}.  Figure \ref{fig3}(lower panel) presents
the time evolution of populations calculated numerically by
solving the Schr\"{o}dinger equation.

Figure \ref{fig4} represents the reverse pulse  sequence of
cavity-Stark which corresponds to the path (b) in Fig. \ref{fig2}.
We see that for the two sequences of the pulses, the population is
completely transferred from the initial state $|+,0\rangle$ to the
target state $|-,1\rangle$, i.e., the generation of a
single-photon Fock state in the cavity mode  with atomic
population transfer to the ground state $|-\rangle$ at the end of
the SCRAP process.

%This SCRAP process is more robust with respect to variations of
%the cavity Rabi frequency than with respect to variations of the
%Stark shift. The reason comes back to the special structure of the
%eigenenergy surfaces in Fig. \ref{fig2}. We can see that on the
%$S=0$ plane, between the two surfaces there is not any
%intersection, while on the plane $G=0$ between these surfaces a
%conical intersection is present. In general, as the number of
%conical intersections between neighboring eigenenergy surfaces
%increases, the robustness of the adiabatic passage is decreased.

It can be shown \cite{YatsenkoPRA99} that for time-delayed
Gaussian-shaped pulses (\ref{Grabi}) and (\ref{Srabi}), the
condition of adiabaticity translates to the following requirements
for the experimentally controllable pulse parameters
$G_{0},~T_{C},~\delta,~S_{0},~T_{S},$ and $\tau$,
\begin{equation}\label{ad-cond}
\exp\left(\frac{-8\tau^{2}}{T_{C}^{2}}\right)\ll\frac{\delta}{G_{0}^{2}T_{S}}\sqrt{\ln\frac{S_{0}}{\delta}}\ll
1.
\end{equation}
The inequalities (\ref{ad-cond}) imply that it is preferable to
work with large cavity-field amplitudes and small static detuning.

The  typical value of the cavity lifetime is of
 the order of $T_{\text{cav}}=1$ ms corresponding to $Q=3\times
 10^{8}$, and the  upper limit of interaction time is  $T_{S}\sim T_{C}\sim T_{\text{int}}=100$ $\mu$s (atom with a velocity
 of 100 m/s with the cavity mode waist of $W_{C}=6$ mm)
 \cite{raimond}. The condition of global adiabaticity
$G_{0}~T_{\text{int}}\gg
 1$ for the typical value of $G_{0}\approx 0.15$ MHz \cite{raimond} is
 well satisfied $(G_{0}~T_{\text{int}}\approx 15$).
 Numerics (Figs. \ref{fig3},\ref{fig4} shows additionally that the diabatic
 dynamics through the conical intersections of Fig. \ref{fig2} also
 satisfied. Closed cavities with  higher $Q$ factor $Q=4\times 10^{10}$ and
longer decay time $T_{\text{cav}}=0.3$
 s may also be used \cite{walther}.

The adiabatic passage can be optimized for a dynamics following a
trajectory close to a level line of the difference of the two
surfaces of Fig. \ref{fig2}, corresponding to parallel
instantaneous eigenenergies \cite{guerinPRA2002,guerinACP}.

One issue that needs to be addressed in more detail, is the role
of cavity damping which results from the population of the state
$|-,1\rangle$ during the time evolution of the system. The cavity
damping induce loss that goes from the state $|-,1\rangle$ to the
state $|-,0\rangle$, which is not  coupled by the effective
Hamiltonian. We make the simulation with a loss term in the
Schr\"{o}dinger equation for the state $|-,1\rangle$. This term is
$(-i/T_{\text{cav}})$  on the diagonal of the effective
Hamiltonian (\ref{Heff}) associated to the state $|-,1\rangle$.

\begin{figure}
\includegraphics[width=6cm]{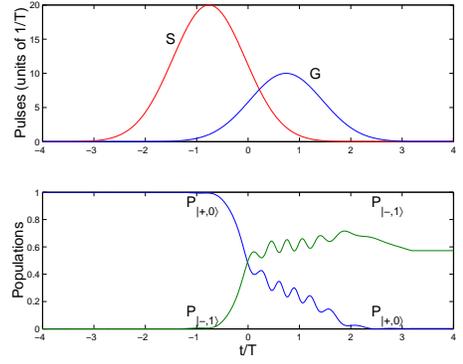}
 \caption{(Color online). Same as Fig. \ref{fig3},
 but with inclusion of cavity damping in the effective Hamiltonian as
 $T_{\text{cav}}=10~T_{\text{int}}$. The lower panel shows that the final state of the system will be
$|-,1\rangle$ with the population of 0.57.}\label{damp10}
\end{figure}

\begin{figure}
\includegraphics[width=6cm]{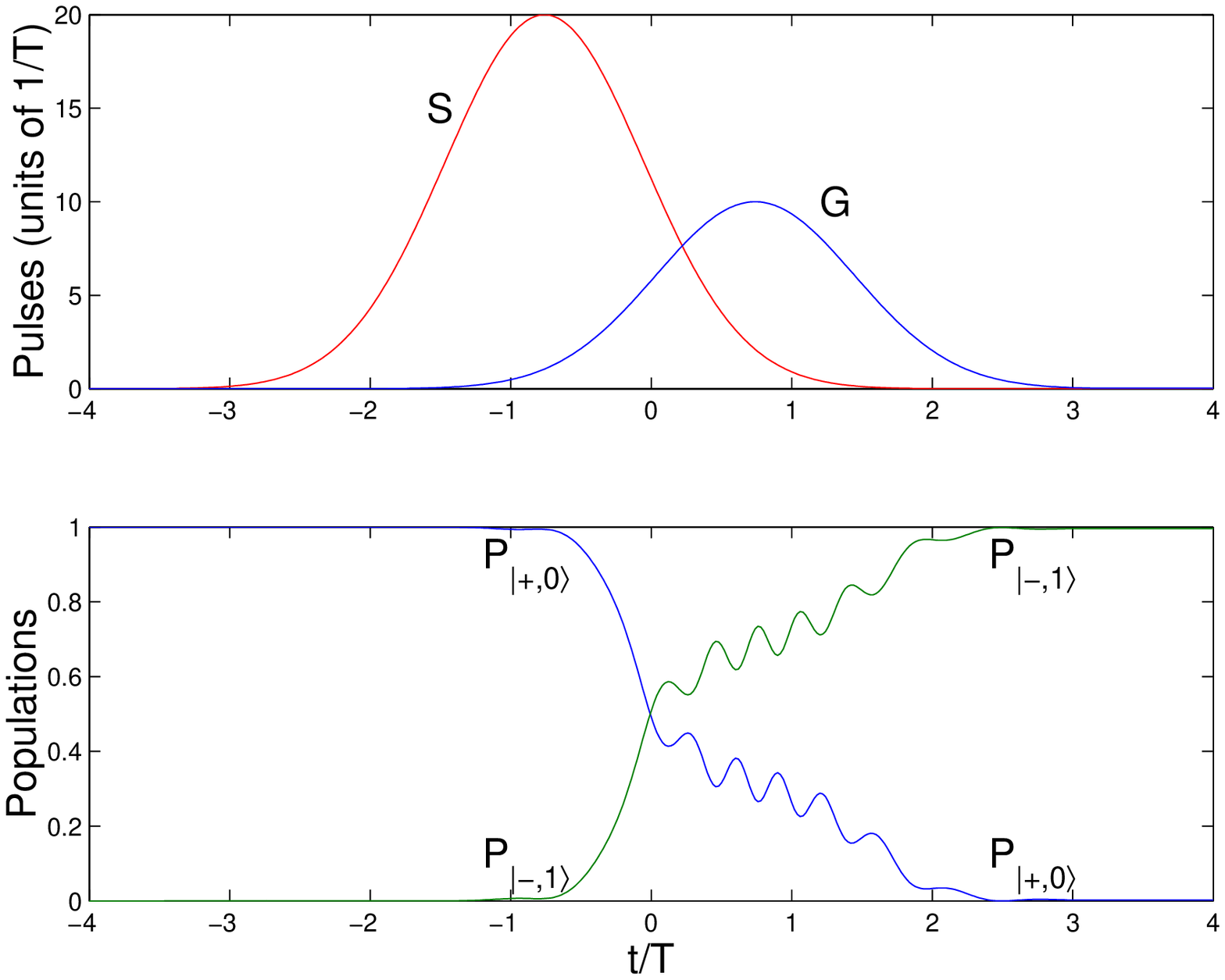}
 \caption{(Color online). Same as Fig. \ref{fig3},
 but with inclusion of cavity damping in effective Hamiltonian as
 $T_{\text{cav}}=3000~T_{\text{int}}$. The lower panel shows that the final state of the system will be
$|-,1\rangle$ with the population of 0.99.}\label{damp3000}
\end{figure}

Figure \ref{damp10} (resp. \ref{damp3000}) shows that taking into
account the cavity damping leads to a final population 0.57 (resp.
0.99) of the state $|-,1\rangle$ for $T_{\text
{cav}}=10~T_{\text{int}}$ (resp. $T_{\text{cav}}=3000~T_{\text{
int}}$), which for $T_{\text{int}}=100~\mu$s corresponds to
$T_{\text{cav}}=1$ ms (resp. $T_{\text{cav}}=0.3$ s).
%%%%%%%%%%%%%%%%%%%%%%%%%%%%%%%%%%%%%%%%%%%%%%%%%%%%%%%%%%%%%%%%%%%%%%%%%%%%%
%%%%%%%%%%%%%%%%%%%%%%%%%%%%%%%%%%%%%%%%%%%%%%%%%%%%%%%%%%%%%%%%%%%%%%%%%%%%%%%

%%%%%%%%%%%%%%%%%%%%%%%%%%%%%%%%%%%%%%%%%%%%%%%%%%%%%%%%%%%%%%%%%%%%%%%%%%%%%
%%%%%%%%%%%%%%%%%%%%%%%%%%%%%%%%%%%%%%%%%%%%%%%%%%%%%%%%%%%%%%%%%%%%%%%%%%%%%%%

\section{\label{ape}atom-photon entanglement}
%%%%%%%%%%%%%%%%%%%%%%%%%%%%%%%%%%%%%%%%%%%%%%%%%%%%%%%%%%%%%%%%%%%%%%%%%%%%%%%%%%%%%%%%%%%%%%%%%%%
\begin{figure}
  \includegraphics[width=6cm]{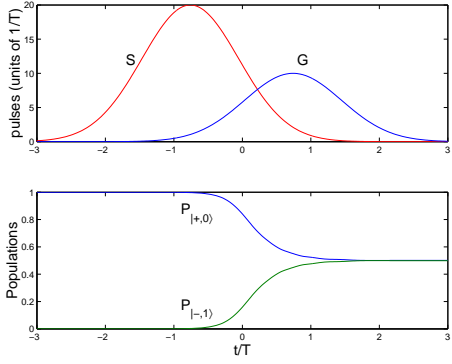}\\
  \caption{(Color online). Upper panel:  Sequence of the Stark and cavity (Gaussian-shaped)
  pulses with the same parameters of Fig. \ref{fig3} and $\delta=0$. Lower panel: Populations as function of time which represent
  a transfer of population
   from the state $|+,0\rangle$ to a superposition of $|+,0\rangle$ and $|-,1\rangle$ with equal weights
   (maximally atom-photon entangled state) at the end of the half-SCRAP process.}\label{fig5}
\end{figure}
%%%%%%%%%%%%%%%%%%%%%%%%%%%%%%%%%%%%%%%%%%%%%%%%%%%%%%%%%%%%%%%%%%%%%%%%%%%%%%
For a certain range of the cavity detuning $\delta$, the SCRAP
technique acts as a fractional SCRAP and will produce a coherent
superposition of the states $|+,0\rangle$ and $|-,1\rangle$. The
composition of the created superposition is controlled by $\delta$
and is robust against variations in the other interaction
parameters. Half SCRAP is a variation of SCRAP for which the
cavity detuning  $\delta$ is zero, and thus  $\Delta=S=-|S|$. Both
pulse sequences of cavity-Stark  and Stark-cavity leads to the
same populations of the states $|+,0\rangle$ and $|-,1\rangle$ at
the end of the process. This corresponds to the generation of
maximally atom-photon entangled states
$\frac{1}{\sqrt{2}}(|+,0\rangle+|-,1\rangle)$. However the
sequence of cavity-Stark is not interesting from the view point of
applications, since it results in the superposition an additional
dynamical phase factor which is difficult to control in a real
experiment \cite{YatsenkoOC02}. Hence, we consider in the
following only the pulse sequence of Stark-cavity.

The instantaneous eigenstates of the effective Hamiltonian
(\ref{Heff}), or adiabatic states of the system, are as follows:
\begin{eqnarray}\label{ADbasis}
|\phi_{-}\rangle&=&\cos \vartheta(t) |+,0\rangle- \sin\vartheta(t)
|-,1\rangle,\nonumber\\
 |\phi_{+}\rangle&=&\sin \vartheta(t)|+,0\rangle+ \cos\vartheta(t)|-,1\rangle,
\end{eqnarray}
where the mixing angle $\vartheta(t)$ is defined  as
\begin{equation}\label{mix-angl}
    \tan 2\vartheta=\frac{G}{\Delta},\qquad 0\leq\vartheta<\pi/2.
\end{equation}
When the Stark pulse precedes the cavity pulse, we have:
\begin{equation}\label{hscrap-lim}
\lim_{t\rightarrow t_{i}}\frac{G}{-|S|}=0,\qquad
\lim_{t\rightarrow t_{f}}\frac{G}{-|S|}=-\infty.
\end{equation}
This leads to $\vartheta(t_{i})=0$ and $\vartheta(t_{f})=-\pi/4$,
and
\begin{subequations}
\begin{eqnarray}\label{hscrap-state-lim}
    |+,0\rangle\xleftarrow[]{t_{i}\leftarrow t}&|\phi_{-}(t)\rangle&\xrightarrow[]{t\rightarrow
    t_{f}}\frac{1}{\sqrt{2}}(|+,0\rangle+|-,1\rangle),\\
|-,1\rangle\xleftarrow[]{t_{i}\leftarrow
t}&|\phi_{+}(t)\rangle&\xrightarrow[]{t\rightarrow
    t_{f}}\frac{1}{\sqrt{2}}(-|+,0\rangle+|-,1\rangle).
\end{eqnarray}
\end{subequations}
Hence if the initial state of the system is taken as $|\Psi(t_{i})
\rangle=|+,0\rangle$, the state $|\phi_{-}(t)\rangle$ is the only
adiabatic state populated during the whole evolution in the
adiabatic limit, and no population will transfer to the other
adiabatic state $|\phi_{+}(t)\rangle$. Hence, the system will be
driven from the state $|+,0\rangle$ into the superposition
\begin{equation}\label{hscrap-Psi-fin}
|\Psi(t_{f})
\rangle=\frac{1}{\sqrt{2}}\left(|+,0\rangle+|-,1\rangle\right),
\end{equation}
except for an irrelevant common phase factor.

Figure \ref{fig5} (upper panel) represents the  pulse sequence of
Stark-cavity with the same pulse parameters of Fig. \ref{fig3} and
the necessary condition $\delta=0$. We see in the lower panel of
this figure that  the population is completely transferred from
the initial state $|+,0\rangle$ to the target state
$\frac{1}{\sqrt{2}}(|+,0\rangle+|-,1\rangle)$, i.e., the
generation of a maximally atom-photon entangled state
 at the end of the half-SCRAP process.
 \section{\label{aae}atom-atom entanglement}
 In this section, we show that by combining  half-SCRAP and
SCRAP processes for two traveling atoms, we can prepare the atoms
in a maximally entangled state
$\frac{1}{\sqrt{2}}(|+,-\rangle+|-,+\rangle)$.
 An interesting property of the SCRAP process is that for both of
 the pulse sequences cavity-Stark and Stark-cavity, the initial states
 $|+,0\rangle$ and $|-,1\rangle$ evolve as follows:
 \begin{equation}\label{evolv1}
|+,0\rangle\xrightarrow[]{\text{SCRAP}}|-,1\rangle,\qquad
|-,1\rangle\xrightarrow[]{\text{SCRAP}}|+,0\rangle.
\end{equation}
 We also notice that
the initial state  $|-,0\rangle$ evolves under a SCRAP process as
follows:
\begin{equation}\label{evolv2}
|-,0\rangle\xrightarrow[]{\text{SCRAP}}|-,0\rangle.
\end{equation}

In the following we consider the state of the atom1-atom2-cavity
system as $|A1,A2,n\rangle$ where $\{A1,A2=+,-\}$, and $\{n=0,1\}$
is the number of photons in the cavity-mode. To create atom-atom
maximal entanglement in a microwave cavity, we suppose that the
initial state of the combined system is $|+,-,0\rangle$. Two atoms
enter successively the cavity mode with different static detunings
$\delta$. The first atom encounters the pulse sequence of
Stark-cavity in the frame of half-SCRAP process. At the second
step, the second atom encounters the same pulse sequence in the
frame of SCRAP process. The quantum state of the combined system
evolves as follows:
\begin{eqnarray}
  |+,-,0\rangle &\xrightarrow[]{\text{half-SCRAP}}& \frac{1}{\sqrt{2}}
(|+,-,0\rangle+|-,-,1\rangle)\nonumber\\
   &\xrightarrow[]{\text{SCRAP}}&\frac{1}{\sqrt{2}}
(|+,-,0\rangle+|-,+,0\rangle),
\end{eqnarray}
which corresponds to a  pair of atoms in a maximally entangled
atomic state with an empty cavity. The cavity field which starts
and ends up in the vacuum state and remains at the end of the
process uncorrelated from the atoms, acts as a \emph{catalyst} for
the atom-atom entanglement.

\section{\label{QST} quantum state transfer and quantum networking}
In this section we use the SCRAP technique to show that: (i) an
unknown state of a two-level atom,
$\alpha|-\rangle+\beta|+\rangle$ where $\alpha$ and $\beta$ are
unknown arbitrary coefficients, can be transferred to another one
with the initial state of $|-\rangle$ through a microwave cavity
mode which is initially in the state $|0\rangle$; (ii) an unknown
state of a cavity mode $\alpha|0\rangle+\beta|1\rangle$ can be
transferred to another cavity (initially in the  state
$|0\rangle$) through a two-level atom (initially in the  state
$|-\rangle$).

\begin{figure}
  \includegraphics[width=8cm]{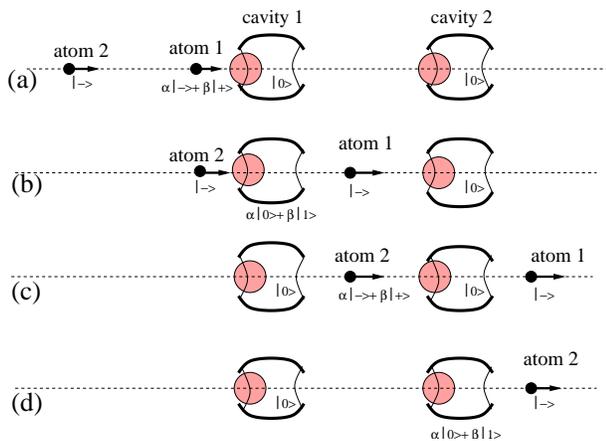}\\
  \caption{(Color online).  Schematic diagram for the QST protocol between two identical atoms
  via a cavity mode, and for quantum network between two distant cavities
  via the atomic channel.}\label{qst}
\end{figure}

Figures \ref{qst}-a,b,c   demonstrates how the unknown state
$(\alpha|-\rangle+\beta|+\rangle)$ of atom 1 is transferred to
atom 2 through the cavity 1 (initially in the  state $|0\rangle$).
We send the atom 1 through the cavity 1 in the frame of SCRAP
process. Equations (\ref{evolv1}) and (\ref{evolv2}) shows that
the final states of cavity 1 and atom 1 will be
$(\alpha|0\rangle+\beta|1\rangle)$, $|-\rangle$ at the end of
SCRAP process. After atom 1 comes out of the cavity 1, atom 2
(initially in the state $|-\rangle$) is sent through the cavity 1
(initially in the state $\alpha|0\rangle+\beta|1\rangle$). Atom 2
also experiences  a SCRAP process during the interaction with the
cavity 1. The final states of cavity 1 and atom 2 at the end of
the second SCRAP process will be $|0\rangle$ and
$(\alpha|-\rangle+\beta|+\rangle)$ respectively. This means that
not only the unknown state $(\alpha|-\rangle+\beta|+\rangle)$ is
transferred from atom 1 to atom 2, but also the atoms exchange
their states via  an interaction with a microwave cavity mode.

Figures \ref{qst}-b,c,d   demonstrates how the unknown state
 of the cavity 1 is transferred
to the cavity 2 (initially in the  state $|0\rangle$) by SCRAP
technique. We show that we can prepare a quantum network, in which
long-lived atomic states (quantum channel) are used to communicate
between the two nodes of the network. We assume that there are two
identical cavities 1 and 2, which are considered as two nodes of
the network. Now our goal is to transfer the state
($\alpha|0\rangle+\beta|1\rangle$) of cavity 1 (node 1) to the
cavity 2 (node 2). For that we send an atom (atom 2 in Fig.
\ref{qst}) through the cavity 1. We see that this atom is prepared
in the state $(\alpha|-\rangle+\beta|+\rangle)$ at the end of
interaction with the node 1. This atom is now sent through the
cavity 2 (node 2) which is initially in the state $|0\rangle$. In
this way, the unknown state ($\alpha|0\rangle+\beta|1\rangle$) of
node 1 is transferred to the node 2. This idea can be extended to
a number of distant nodes. For example, to send the state
($\alpha|0\rangle+\beta|1\rangle$) to a third cavity (node 3), we
can send a third atom in the same direction after the second one
in Fig. \ref{qst}.

%%%%%%%%%%%%%%%%%%%%%%%%%%%%%%%%%%%%%%%%%%%%%%%%%%%%%%%%%%%%%%%%%%%%%%%%%%%%%%%%%%%%%%%%%%%%%%%%%%%
%%%%%%%%%%%%%%%%%%%%%%%%%%%%%%%%%%%%%%%%%%%%%%%%%%%%%%%%%%%%%%%%%%%%%%%%%%%%%%

\section{\label{con}discussions and conclusions}
Using the topological properties of  eigenenergy surfaces of the
effective Hamiltonian of the atom-cavity system and SCRAP and
half-SCRAP techniques, we have engineered single-photon Fock state
and maximally atom-photon and atom-atom entangled states. We also
proposed a deterministic way, using SCRAP technique, for quantum
networking. This protocol does not require any kind of probability
arguments based on the outcome of a measurement. The realization
of parameters satisfying the conditions of the proposed schemes
appears feasible with progressive improvements to experiments with
high-Q microwave cavities. In this analysis we have assumed that
the interaction time between the two-state atom and the fields is
short compared to the
 cavity lifetime $T_{\text{cav}}$ and the atom's excited state lifetime
 $T_{\text{at}}$, i.e. $T_{\text{int}}\ll T_{\text{cav}} , T_{\text{at}}$, which are essential for an
 experimental setup and avoiding  decoherence effects.

The tunable lasers allows the excitation of highly excited atomic
states, called Rydberg states. Such excited atoms are very
suitable for observing quantum effects in radiation-matter
coupling for two reasons: First, these states are very strongly
coupled to the radiation field; Furthermore,
 Rydberg states have relatively long lifetimes with
respect to spontaneous decay. In the microwave domain, the
 radiative lifetime of circular Rydberg states -- of the order of $T_{\text{at}}=30
 $ ms -- are much longer than those for non-circular  Rydberg states.
 %%%%%%%%%%%%%%%%%%%%%%%%%%%%%%%%%%%%%%%%%%%%%%%%%%%%%%%%%%%%%%%%
\begin{acknowledgments}
M. A-T. wishes to acknowledge the financial support of the MSRT of
Iran and SFERE.
\end{acknowledgments}
%%%%%%%%%%%%%%%%%%%%%%%%%%%%%%%%%%%%%%%%%%%%%%%%%%%%%%%%%
\bibliography{paper5}
%%%%%%%%%%%%%%%%%%%%%%%%%%%%%%%%%%%%%%%%%%%%%%%%%%%%%%%%%
\end{document}